\documentclass[pre,twocolumn,showpacs,floatfix,superscriptaddress]{revtex4}

\usepackage{graphicx}
\usepackage{dcolumn}
\usepackage{amsmath}
\usepackage{bm}
\begin{document}

\title{Elasticity of polyelectrolyte multilayer microcapsules}

\author{Valentin V. Lulevich}
\affiliation{Max Planck Institute for Polymer Research, Ackermannweg 10, 55128 Mainz, Germany}
\affiliation{Laboratory of Physical Chemistry of Modified Surfaces, Institute of Physical Chemistry, Russian Academy of Sciences, 31 Leninsky Prospect, 119991 Moscow, Russia}

\author{Denis Andrienko}
\affiliation{Max Planck Institute for Polymer Research, Ackermannweg 10, 55128 Mainz, Germany}

\author{Olga I. Vinogradova}
\email[Corresponding author: ]{vinograd@mpip-mainz.mpg.de}
\affiliation{Max Planck Institute for Polymer Research, Ackermannweg 10, 55128 Mainz, Germany}
\affiliation{Laboratory of Physical Chemistry of Modified Surfaces, Institute of Physical Chemistry, Russian Academy of Sciences, 31 Leninsky Prospect, 119991 Moscow, Russia}

%\date{\today}

\begin{abstract}
  We present a novel approach to probe elastic properties of
  polyelectrolyte multilayer microcapsules. The method is based on
  measurements of the capsule load-deformation curves with the atomic
  force microscope. The experiment suggests that at low applied load
  deformations of the capsule shell are elastic. Using elastic theory
  of membranes we relate force, deformation, elastic moduli, and
  characteristic sizes of the capsule. Fitting to the prediction of
  the model yields the lower limit for Young's modulus of the
  polyelectrolyte multilayers of the order of $1-100$ MPa,
  depending on the template and solvent used for its dissolution.
  These values correspond to Young's modulus of an elastomer.
\end{abstract}
\maketitle

\section{Introduction}

Recently there has been much interest in studying of the polyectrolyte
multilayer microcapsules~\cite{donath.e:1998}. These capsules are made
by layer-by-layer (LbL) adsorption~\cite{decher.g:1997} of oppositely
charged polyelectrolytes onto charged colloidal particles with
subsequent removal of the template core~\cite{sukhorukov.gb:1998.b}
and are important for a variety of applications such as drug delivery,
catalysis, and biotechnology. At a more fundamental level,
polyelectrolyte microcapsules represent a convenient system for
studying physical properties of free polyelectrolyte multilayer films.
A free thin film geometry allows one to study properties not
accessible in the bulk or in the supported films. In this way one
might gain a better understanding of polyelectrolytes in general.

Mechanical properties are among the most important and almost
certainly the least understood physical properties of free multilayer
films.  They define deformation and rupture of the capsule shell under
an external load, which is important 
for instance in the development of the delivery systems for drug
injection into blood vessels \cite{lewis.dd:1990}. This can also help
to understand the principles underlying the mechanical behavior of
living cells. For example, elastic properties of the capsule shell can
be used as an input to models of penetration of particular viruses
into the biological cell~\cite{deserno.m:2002.a}.

There have been three recent attempts to study mechanical behavior of
polyelectrolyte microcapsules.  The first involves observing
osmotically induced buckling of capsules immersed in a polyelectrolyte
solution~\cite{gao.c:2001}. In this method the external osmotic
pressure induces instability in the capsule shape. If the capsule is
highly permeable for the solvent (water) the capsule wall bends and a
cup-like shape is formed. The second method is based on studying the
swelling of microcapsules filled with the solution of a strong
polyelectrolyte~\cite{vinogradova.oi:2003}. The third attempt involves
measuring the deformation of microcapsules under applied load using
atomic force microscope
(AFM)~\cite{lulevich.vv:2002,lulevich.vv:2003}. The advantage of the
AFM technique is its accuracy, the possibility of studying a wider
range of systems, and richer experimental information: it allows one
to distinguish between different regimes in load-deformation profiles,
study permeability, elasticity, and plasticity of the capsule shell.

Motivated by a recent AFM
study~\cite{lulevich.vv:2002,lulevich.vv:2003}, which was focused on
the {\em qualitative} difference between various regimes in the
load-deformation profiles and on the role of encapsulated polymers, we
apply the AFM approach to study {\em quantitatively} the elastic
properties of the capsule shell. As an initial application of our
method we have chosen to study the capsules with the shells composed
of layers of alternating poly(sodium 4-styrene sulfonate) and
poly(allylamine) hydrochloride.  These multilayers are stable in
various chemical environments. Their thickness is known to grow
linearly with the the number of deposited bilayers, so that they
represent a convenient system for verification of our approach. Beside
that this type of the shells was used in the previous studies of
mechanical properties of the
capsules~\cite{gao.c:2001,vinogradova.oi:2003,lulevich.vv:2002,lulevich.vv:2003}.
We work in a small deformation regime, which allows one to treat
capsule deformations as elastic. By fitting load-deformation curves to
the predictions of a simple model we are able to obtain the lower
limit for Young's modulus of the polyelectrolyte multilayer complex
forming the capsule shell.

\section{Materials and Methods}

The capsules were produced according to the method~\cite{sukhorukov.gb:1998.b,shenoy.db:2003} based on LbL assembly of several bilayers of polystyrene sulfonate (PSS, ${\rm Mw} \sim 70000$, Aldrich) and poly(allylamine) hydrochloride (PAH, Mw $\sim 50000$, Aldrich) layers on a weakly crosslinked (latex) template. As a template we used suspensions of polydisperse Poly-DL-lactic acid (PLA) particles and monodisperse melamine formaldehyde particles (MF) purchased from Microparticles GmbH (Berlin, Germany). The PLA particles were of a radius in the range from 2 to 8 $\mu$m and were dissolved using 1:1 mixture of (1-methyl-2-pyrrolindinone) acid  and organic solvent (acetone) after assembly of four PSS/PAH bilayers~\cite{shenoy.db:2003}. This type of capsule is referred below as PLA capsules. The MF particles were of radius $2.0 \pm 0.1$  $\mu$m and were dissolved using HCl (at pH $1.2-1.4$) to give so-called MF capsules. The variation of shell thickness was achieved by two methods. In the first method we varied the number of bilayers deposited on MF templates. In the second method we dissolved the MF particles after the deposition of the fourth bilayer. Then several additional bilayers were deposited to tune the final shell thickness.

Load (force) vs deformation curves were measured with the Molecular Force Probe (MFP) 2D AFM (Asylum Co, Santa Barbara, USA) equipped with a nanopositioning sensor that corrects piezoceramic hysteresis and creep. The MFP was used together with a commercial confocal microscope manufactured by Olympus (Japan) consisting of the  conconfocal laser scanning unit Olympus FV 300 in combination with 
an inverted fluorescence microscope Olympus IX70 equipped with a high-resolution immersion oil objective (60$\times )$. The confocal microscope was specially adjusted for the MFP 2D. This allowed simultaneous optical measurements of the capsule's shape during the AFM force experiment. The excitation wavelength was chosen according to the label Rhodamine (543 nm). The z-position scanning was done in steps of 0.05 $\mu$m. 

\begin{figure}
\begin{center}
\includegraphics[width=6cm]{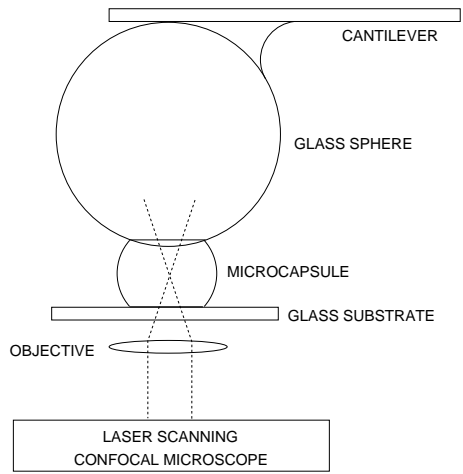}
\end{center}
\caption[]{% fig1.eps
  Schematic of the AFM force experiment.} 
  \label{fig:setup}
\end{figure}

A schematic of the experiment is presented in Fig.~\ref{fig:setup}. A drop (50-100 $\mu \rm l$) of water (purified by a commercial Milli-Q Gradient A10 system containing ion exchange and charcoal stages) suspension of polyelectrolyte microcapsules was deposited onto a thin glass slide fixed over the oil immersion objective of the confocal microscope. The glass sphere (radius $R_{\rm s} = 20 \pm 1 \mu \rm m$, Duke Sci. Co., California) attached to a cantilever (V-shaped, Micromash, Estonia, spring constants $k = 2.5 {\rm N/m}$) was centered above the apex of a capsule with accuracy of $\pm 0.5 \mu \rm m$, using the graticule lines in the optical image for alignment. Measurements were performed at a speed in the range from 0.2 to 20 $\mu \rm m/s$.

The result of measurements represents the deflection $\Delta$ vs the position of the piezo translator at a single approach (loading). The load, $\cal F$, was determined from the cantilever deflection, ${\cal F} = k \Delta$. As before, we assume that the zero of separation is at the point of the first measurable force ~\cite{lulevich.vv:2002}. Then the deformation is calculated as the difference between the position of the piezotranslator and cantilever deflection.

\begin{figure}
\begin{center}
\includegraphics[width=5cm]{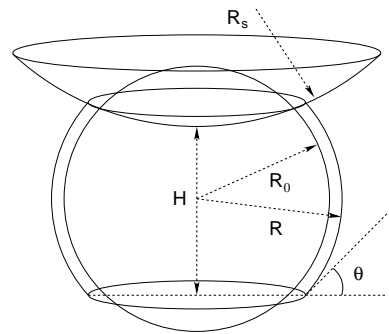}
\end{center}
\caption[]{% fig1.eps
  Sketch of the microcapsule: undeformed and deformed states. If the
  volume of the deformed capsule remains constant (no water drainage
  through the capsule shell), then the increase in the capsule radius
  is quadratic in the relative deformation $\epsilon = 1 - H/(2R_0)$
  (see Eq.~(\ref{eq:radius})).  } 
  \label{fig:geometry}
\end{figure}

The diameters of the capsules were determined optically with the accuracy of $0.4$ $\mu$m and from the AFM load vs deformation curves (see~\cite{lulevich.vv:2002} for more details). The relative deformation $\epsilon$ of the capsule was then defined as $\epsilon = 1 - H/(2R_0)$ (see Fig.~\ref{fig:geometry}), where $R_0$ is the radius of the undeformed capsule.

\section{Results}

\subsection{MF capsules}
\begin{figure}
\begin{center}
\includegraphics[width=8cm]{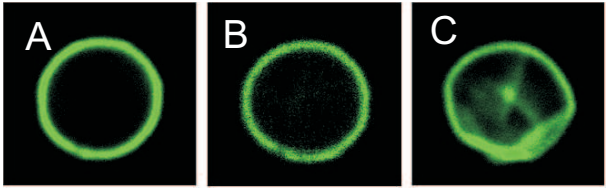}
\end{center}
\caption[]{% fig1.eps
  Typical confocal images of MF capsules (radius $R_0 = 2\mu {\rm m}$)
  obtained for different stages of deformation: (a) $\epsilon
  = 0$, (b) $\epsilon = 0.2$, and (c) $\epsilon = 0.5$. }
  \label{fig:image}
\end{figure}

The 3D confocal scanning suggested that undeformed capsules have spherical shape (Fig.~\ref{fig:image}a).

\begin{figure}
\begin{center}
\includegraphics[width=8cm]{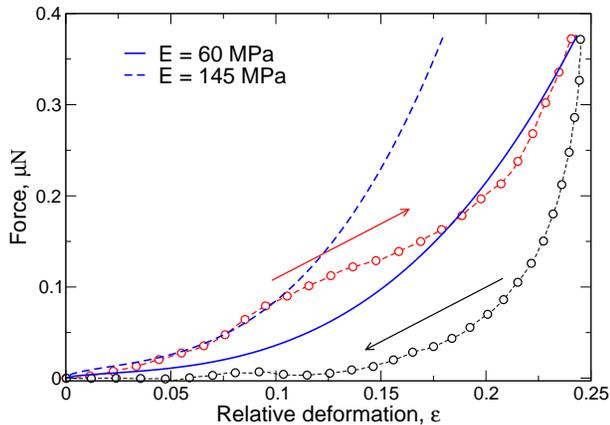}
\end{center}
\caption[] {% fig3.eps
  Typical loading/unloading curves for MF capsules (radius $R_0 =
  2\mu {\rm m}$). Driving speed $v = 20 \mu \rm {m / s}$. Fit to the
  Eq.~(\ref{eq:force}) gives an estimate for the Young's modulus in
  the range $E \approx 60 - 150 {\rm MPa}$.}
  \label{fig:hysteresis}
\end{figure}

Fig.~\ref{fig:hysteresis} shows typical for MF capsules
(loading/unloading) deformation profiles for small relative deformations ($\epsilon
\lesssim 0.2 - 0.3$). By analyzing these profiles and images of
confocal scanning we found that:

(i) The deformation is completely reversible. In other words, the
capsule always returns to its original state (the force vanishes at
zero deformation) during consequent loading/unloading cycles. No deviation
of the free area of the capsules from the spherical shape was found
within the accuracy of confocal measurements (Figs.~\ref{fig:image}a,b);

(ii) The pull-off force is equal to zero; 

(iii) The load-deformation profiles do not depend on the driving speed; 

(iv) The loading/unloading has hysteresis, i.e. force upon retraction is less than on approach at the same relative deformation. This loading/unloading hysteresis is relatively small.

For relative deformations $\epsilon \gtrsim  0.2 - 0.3$ we observed large hysteresis, only partial reversibility in loading/unloading, noisy regions in the deformation profiles, and significant dependence on the driving speed, which indicates drainage of the inner solution through the shell and/or its local ruptures. This is also confirmed by a confocal 3D scanning (Fig.~\ref{fig:image}c).

\begin{figure}
\begin{center}
\includegraphics[width=8cm]{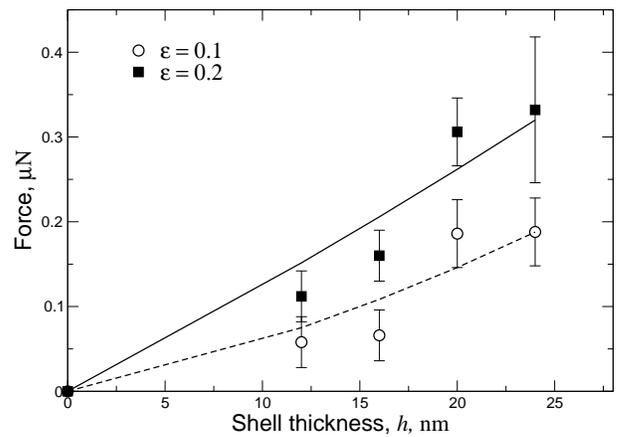}
\end{center}
\caption[] {% fig3.eps
  Load vs shell thickness for MF capsules of $R_0 = 2 \mu \rm m$.
  Circles: relative deformation $\epsilon = 0.1$, squares: $\epsilon =
  0.2$.  Fit to Eq.~(\ref{eq:force}) gives $E \approx 60 {\rm MPa}$
  for $\epsilon = 0.2$, and $E \approx 180 {\rm MPa}$ for $\epsilon =
  0.1$.  Measurements were performed at the driving speed $20\mu \rm
  m/s$.  }
  \label{fig:thickness}
\end{figure}

The load at a fixed relative deformation for MF capsules with different
shell thickness $h$ is presented in Fig.~\ref{fig:thickness}. The value of $h$ was calculated as a product of the number of PSS/PAH bilayers in the shell and a thickness of one bilayer. The values reported for a bilayer thickness vary in the range $\approx 3-5$ nm~\cite{sukhorukov.gb:1998.a,sukhorukov.gb:1998.b}. Here we use the average value of 4 nm. Analysis of data presented in Fig.~\ref{fig:thickness} suggests that the
capsules are getting stiffer with the increase in the shell thickness, and that the effect of the shell thickness depends on the relative deformation 
  $\epsilon$.

\subsection{PLA capsules}

By performing the same force and optical measurements we have made exactly the same conclusions as for MF capsules. PLA capsules have, however, been found to be much softer (weaker force at the same $\epsilon$), with much smaller hysteresis, i.e. the loading branch of the force was much closer to the unloading one. 

\begin{figure}
\begin{center}
\includegraphics[width=8cm]{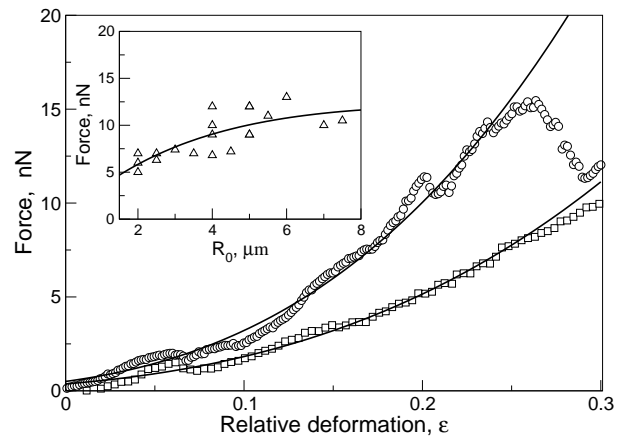}
\end{center}
\caption[] {% fig3.eps
Load vs deformation curves for PLA capsules of different size (symbols) and their fit to Eq.~(\ref{eq:force}) (solid curves). Squares: $R_0 = 2 \mu \rm m$, circles: $R_0 = 5\mu \rm m$. Inset shows the force at a fixed relative deformation, $\epsilon = 0.2$, as a function of the capsule radius. Measurements were performed at the driving speed $2 \mu \rm m/s$. 
} \label{fig:load}
\end{figure}

With the PLA capsules we were able to investigate the effect of the capsule size. Analysis of the load-deformation curves (Fig.~\ref{fig:load}) showed that bigger capsules are stiffer than smaller ones. The inset of Fig.~\ref{fig:load} shows the force at a fixed relative deformation as a function of the capsule radius. Each point corresponds to a different capsule. The scatter in the force is due to the variation of the properties of the capsule shell, which changes slightly from capsule to capsule (the actual error of our force measurements was about $0.5 \rm nN$).

\section{Discussion}
%\subsection{Theoretical model}
A simple model accounts for these experimental observations. Since
hysteresis is small and the load-deformation profiles do not depend on
the driving speed (in the interval used) we can, in the first
approximation, neglect drainage of the water through the capsule shell
and assume that the volume of the capsule does not change on short
time scale of the AFM experiment.

To relate the relative deformation $\epsilon$ and the change in the
capsule radius we further assume that, for small deformations, the
shape of the capsule remains spherical except in the contact
regions~\cite{shanahan.mer:1997} (see Fig.~\ref{fig:geometry}).  Then,
for small $\epsilon$, volume conservation requires
\begin{equation}
R = R_0+\frac{R_0}{2}\frac{1+R_0/(2R_{\rm s})}{(1+R_0/R_{\rm s})^2} \epsilon^2
 +\mathcal{O}(\epsilon^3),
\label{eq:radius}
\end{equation}
where $R$ is the final radius of the capsule. Eq.~(\ref{eq:radius}) shows
that the increase in the capsule radius is quadratic in the relative
deformation $\epsilon$.

Note that there are two dimensionless parameters in
Eq.~(\ref{eq:radius}). The first one is the relative deformation,
$\epsilon$. Deriving Eq.~(\ref{eq:radius}) we expanded $R-R_0$ in
powers of $\epsilon$ and left only the lowest, quadratic in
$\epsilon$, term of the expansion. The second parameter is the ratio
of the glass sphere and the capsule radii, $R_0/R_s$. No expansion was
made with respect to this ratio.  However, when deriving
Eq.~(\ref{eq:radius}), we implicitly assumed that the radii of the
contact areas of the glass sphere and glass substrate are equal, which
is justified provided $R_0$ is at least several times smaller than
$R_{\rm s}$.

%\subsection{Stretching}
Since capsule deformations are small and reversible, we can use the
elastic theory of membranes to find the restoring force.  The elastic
energy of the stretching of the membrane reads ~\cite{landau.ld:1995}
\begin{equation}
G = \frac{h}{2}\int u_{\alpha \beta} \sigma_{\alpha \beta} dS,
\label{eq:membrane_energy}
\end{equation}
where the Einstein summation convention is implied,  
integration is over the membrane's surface, $u_{\alpha \beta}$ is
two-dimensional deformation tensor, and $\sigma_{\alpha \beta}$ is
two-dimensional stress tensor
\begin{equation}
\sigma_{\alpha \beta} = \frac{E}{1-\nu^2}\left[ (1-\nu)u_{\alpha \beta} 
+ \nu \delta_{\alpha \beta} u_{\gamma \gamma}\right],
\end{equation}
where $E$ is Young's modulus, and $\nu$ is Poisson's ratio, and summation over repeating indices is assumed.

Balance of stresses defines the pressure $P$ inside the capsule
\begin{equation}
P = \frac{h}{R} \left( \sigma_{\theta \theta} + \sigma_{\phi \phi} \right) \label{eq:pressure}.
\end{equation}
Using the relation between the stress and deformation tensors
\begin{eqnarray}
Eu_{\theta \theta} = \sigma_{\theta \theta} - \nu \sigma_{\phi \phi}, \\
Eu_{\phi \phi} = \sigma_{\phi \phi} - \nu \sigma_{\theta \theta}, 
\end{eqnarray}
as well as the assumption that the deformed capsule has spherical
shape
\begin{equation}
u_{\theta \theta} = u_{\phi \phi} = \frac{u_r}{R} = \frac{R-R_0}{R},
\label{eq:deform_tens}
\end{equation}
we obtain
\begin{equation}
\sigma_{\theta \theta} = \sigma_{\phi \phi} = \frac{E}{1-\nu}\frac{R-R_0}{R}.
\label{eq:stress_tens}
\end{equation}
Substituting Eqs.~(\ref{eq:deform_tens}, \ref{eq:stress_tens}) to
Eq.~(\ref{eq:membrane_energy}) we obtain the elastic energy due to the
capsule stretching
\begin{equation}
G_{\rm s} \approx 4 \pi \frac{E}{1-\nu} h (R-R_0)^2.
\end{equation}

Correspondingly, the reaction force (load)
\begin{equation}
{\cal F}_{\rm s} = - \frac{\partial G_{\rm s}}{\partial H} = 
2 \pi \frac{E}{1-\nu} h
R_0 \frac{(1+R_0/(2R_{\rm s}))^2}{(1+R_0/R_{\rm s})^4}
\epsilon^3 
\label{eq:stretching_force}
\end{equation}
has cubic dependence on the relative deformation $\epsilon$. For small
microcapsules, i.e. when $R_0/R_{\rm s} \ll 1$, the force at fixed
relative deformation is simply proportional to the capsule radius $R_0$.
The alternative (and equivalent) way to obtain the reaction
force is to multiply the pressure $P$ from Eq.~(\ref{eq:pressure})
by the microcapsule-glass sphere contact area.

%\subsection{Bending} 
Note, that here we ignored the contribution to the elastic energy due
to shell bending. This is reasonable, for a very thin membrane,
everywhere except near the line separating the capsule, glass
sphere/substrate, and the exterior (the edge of the contact area).  An
exact treatment of bending effects is rather complicated; the bending
moment could perturb the spherical shape of the membrane near a contact, etc.
However, simple estimates of the bending energy can be made.  The
change from the membrane in contact with the glass sphere to the free
membrane, at a contact angle $\theta$, occurs over a length comparable to the shell thickness, $h$.
Then the local radius of curvature of the membrane near the
separation line is of the order of $\rho \approx h /
\theta$~\cite{landau.ld:1995,shanahan.mer:2003}. The elastic energy
of bending, $G_{\rm b}$, can be estimated from the beam
theory~\cite{shanahan.mer:2003}
\begin{equation}
G_{\rm b} \approx \frac{EI}{2}\frac{h l}{\rho^2},
\label{eq:bending}
\end{equation}
where $I = h^3/12$ is the second moment of area of `beam'
cross-section, $l = 2 \pi R \sin\theta \approx 2 \pi R \theta$ is the
total length of the line separating the capsule, glass and substrate.
Substituting this back to the expression for the bending free energy
(\ref{eq:bending}) and taking into account the volume conservation requirement $R-R_0 \approx R_0 \theta^4$, we obtain
\begin{equation}
G_{\rm b} \approx \frac{\sqrt{2}\pi}{3} E h^2 R_0\epsilon^{3/2}.
\end{equation}
The reaction force due to shell bending reads
\begin{equation}
{\cal F}_{b}= - \frac{\partial G_{\rm b}}{\partial H} = \frac{\pi}{\sqrt{2}} E h^2 \epsilon^{1/2}.
\label{eq:bending_force}
\end{equation}
Comparing the reaction forces due to stretching and bending,
Eqs.~(\ref{eq:stretching_force}) and (\ref{eq:bending_force}), one can conclude that,
for typical experimental values, $h = 20{\rm nm}$, $R_0 = 2\mu{\rm m}$, bending is
negligibly small for relative deformations $\epsilon \geq 0.15$.

The total reaction force  (load) for $R_s \gg R_0$ and $\nu=1/2$ (typical value for elastomer materials~\cite{shanahan.mer:1997})  reads
\begin{equation}
{\cal F} = 
\frac{\pi}{\sqrt{2}} E h^2 \epsilon^{1/2}+
 4 \pi E h R_0\epsilon^{3}.
\label{eq:force}
\end{equation}

We fitted the load-deformation profiles for PLA capsules to
Eq.~(\ref{eq:force}) taking the Young's modulus, $E$, as a fitting
parameter. Only the data for the capsules with $R < 5\mu{\rm m}$ was
used, to insure that $R_s \gg R_0$.  The fitting curves, together with
the experimental data, are shown in Fig.~\ref{fig:load}. The value $E
= 1 \pm 0.5 \rm MPa$ was obtained from fitting.

For sufficiently large relative deformations, $\epsilon \geq 0.15$,
when we can safely neglect shell bending, our model also predicts the
dependence of the reaction force on the capsule radius, given by the
prefactor in Eq.~(\ref{eq:stretching_force}) which depends on $R_0$.
This prediction is indeed confirmed by the experiment, as shown in the
inset of Fig.~\ref{fig:load}.

For MF capsules, prepared in the acid media without adding any solvent or organic acid, we obtained Young's modulus of the shell of the order of $100 \rm{MPa}$. The experimental dependence of the load on the shell thickness is consistent with our model (Fig.~\ref{fig:thickness}) and confirms that bending is important only at very low $\epsilon$. We also analyzed previously obtained data for MF capsules of the same chemical composition, but prepared by dissolution of a template in a more aggressive acid media (3 mol/L HCl)~\cite{lulevich.vv:2002,lulevich.vv:2003}. For these capsules we obtained a Young's modulus of the PSS/PAH shell of the order of 30 MPa, and even a smaller value for the same capsules, but after treatment in organic solvent (water/acetone mixture)~\cite{lulevich.vv:2003}.

\section{Conclusions}

Several remarks can be made in conclusion. 

First, the Young's modulus obtained here is at least one order of magnitude lower than in the experiment on osmotically induced deformation of PSS/PAH capsules made on the MF template \cite{gao.c:2001}. It can be argued that the theory of osmotically induced shape transition (buckling) of the capsules used in \cite{gao.c:2001} ignores finite permeability of the shell membrane to water. However, it is clear that the finite permeability stabilizes a spherical shape of the capsule, leading to higher values of critical osmotic pressure (infinite for zero permeability) or, equivalently, to smaller values of Young's modulus. The consequences of overestimation of Young's modulus in \cite{gao.c:2001} are very important. It led, in particular, to the conclusion that multilayers represent quite a rigid material comparable with the {\em bulk plastics}~\cite{shackelford.jf:1994}. In contrary to this, our experiment demonstrated that the value of the Young's modulus of the multilayer corresponds to that of {\em elastomers}~\cite{shackelford.jf:1994}. It is likely that the approach (model) we used here underestimates the value of Young's modulus, because it neglects the drainage of water from the capsule. These values ($1-100$ MPa), therefore, give the lowest limit for Young's modulus of polyelectrolyte multilayers. However, our results strongly support the conclusions of the recent swelling experiment~\cite{vinogradova.oi:2003}, which has given a slightly larger, but of the same order of magnitude, value for $E$. We have argued~\cite{vinogradova.oi:2003}, that the osmotically induced swelling measurements give the upper limit for Young's modulus. Therefore, taking together, these two types of experiment provide the range of possible values for Young's modulus of a PSS/PAH multilayer.     

Second, the PLA capsules are one or two order of magnitudes softer than MF capsules~\cite{lulevich.vv:2002}. This is likely due to organic solvent and different (and more concentrated) acid used for a dissolution of the PLA templates. Similar effect is probably responsible for the softening of MF capsules made in different conditions~\cite{lulevich.vv:2002,lulevich.vv:2003}. Thus, our results suggest that the polyelectrolyte multilayers are extremely sensitive to chemical treatment. The effect of softening of the multilayer shell could be due to either structure changes or partial dissolution of the shell, and is poorly understood at the moment~\cite{lulevich.vv:2003}.

Third, the capsules behave as non-Hookean (non-linear) springs even in the weak force regime, contrary to most similar systems which respond as Hookean springs to applied loads \cite{attard.p:2001.a}. 

Forth, the adhesion force does not affect determination of Young's modulus. Indeed, as follows from the model, the area of the contact (and the adhesion energy) is proportional to the relative deformation $\epsilon$. Therefore, the contribution to the force due to adhesion does not depend on $\epsilon$ and will only uniformly shift the load-deformation profiles. 

In summary, using the AFM technique, we were able to estimate the
Young's modulus of the molecularly-thin polyelectrolite multilayer,
which was found to be of the order of a $1-100$ $\rm MPa$. We also
demonstrated that the elastic properties of the multilayer shell can
be modified by treatment in a solvent and acid, and depends on the template used
for the capsule preparation.

\acknowledgments

DA acknowledges the support of the Alexander von Humboldt Foundation. We are grateful to S.~B.~Kim and S.~Nordschild for capsule preparation and to K.~Kremer for helpful discussions.

%\bibliography{paper}

\end{document}